\documentclass[conference]{IEEEtran}
\IEEEoverridecommandlockouts
\usepackage{cite}
\usepackage{amsmath,amssymb,amsfonts}
\usepackage{algorithmic}
\usepackage{graphicx}
\usepackage{textcomp}
\usepackage{xcolor} 
\usepackage{booktabs}
\usepackage{bbding}
\usepackage{multirow}
\usepackage{hyperref}
\hypersetup{
    colorlinks=true,
    linkcolor=black,
    filecolor=magenta,      
    urlcolor=black,
    citecolor = black,
    pdfpagemode=FullScreen,
    }
\input{alphabet}
\definecolor{mypurple}{rgb}{0.8,0.5,0.8}
\definecolor{myviolet}{rgb}{0.6,0.3,0.4}
\definecolor{cadmiumgreen}{rgb}{0.0, 0.42, 0.24}

\newcommand{\DM}[1]{\textcolor{black}{#1}}

\def\Diag{\mathrm{Diag}}

\def\R{\mathbb{R}}
\def\T{^\mathsf {T}}

\def\cc#1{\setlength{\tabcolsep}{0pt}\begin{tabular}{c}#1\end{tabular}}
\newcommand{\figc}[2][]
   {\setlength{\tabcolsep}{0pt}\begin{tabular}{c}\includegraphics[#1]{#2}\end{tabular}}

\begin{document}
\bstctlcite{IEEEexample:BSTcontrol}  
\title{Ultrasound Imaging based on the Variance of a Diffusion Restoration Model\\

\thanks{This work has been supported by the European Regional Development Fund (FEDER), the Pays de la Loire Region on the Connect Talent scheme (MILCOM Project) and Nantes Métropole (Convention 2017-10470).}
}

\author{
\IEEEauthorblockN{Yuxin Zhang\IEEEauthorrefmark{1}, Clément Huneau\IEEEauthorrefmark{1}, Jérôme Idier\IEEEauthorrefmark{1} and Diana Mateus\IEEEauthorrefmark{1}}
\IEEEauthorblockA{\IEEEauthorrefmark{1}Nantes Université, École Centrale Nantes, LS2N,
CNRS, UMR 6004, F-44000 Nantes, France \\
Email: yuxin.zhang@ls2n.fr}
}

\maketitle

\begin{abstract}
Despite today's prevalence of ultrasound imaging in medicine, ultrasound signal-to-noise ratio is still affected by several sources of noise and artefacts. Moreover, enhancing ultrasound image quality involves balancing concurrent factors like contrast, resolution, and speckle preservation. Recently, there has been progress in both model-based and learning-based approaches addressing the problem of ultrasound image reconstruction. Bringing the best from both worlds, we propose a hybrid reconstruction method combining an ultrasound linear direct model with a learning-based prior coming from a generative Denoising Diffusion model. More specifically, we rely on the unsupervised fine-tuning of a pre-trained Denoising Diffusion Restoration Model (DDRM). 
Given the nature of multiplicative noise inherent to ultrasound, this paper proposes an empirical model to characterize the stochasticity of diffusion reconstruction of ultrasound images, and shows the interest of its variance as an echogenicity map estimator. We conduct experiments on synthetic, in-vitro, and in-vivo data, demonstrating the efficacy of our variance imaging approach in achieving high-quality image reconstructions from single plane-wave acquisitions and in comparison to state-of-the-art methods. The code is available at: \href{https://github.com/Yuxin-Zhang-Jasmine/DRUSvar}{https://github.com/Yuxin-Zhang-Jasmine/DRUSvar}.
\end{abstract}


\section{Introduction}
Ultrasound (US) imaging finds extensive use in musculoskeletal, cardiac, obstetrical, and other medical diagnostic applications.
In contrast to Magnetic Resonance or Computer Tomography, which are expensive or ionizing, ultrasound is real-time, affordable, portable, and minimally invasive.
However, US imaging is affected by acoustic attenuation and artefacts (shadowing, reverberation, clutter), as well as electronic and speckle noise.
The standard image reconstruction method, Delay and Sum (DAS)~\cite{DAS}, converts time-domain signals into B-mode images. While this low-complexity approach enables fast reconstruction, it often results in poor image quality in terms of signal-to-noise ratio (SNR), contrast, and spatial resolution, especially with unfocused emissions.

In the last decade, model-based reconstruction methods~\cite{IPB_Ozkan,RED_USIPB} have been proposed, providing physical plausibility but requiring high computational resources for convergence. Meanwhile, deep learning (DL) based techniques~\cite{perdios_cnn-based_2022,van_sloun_deep_2020} have excelled in image denoising and enhancement but face challenges with generalization and interpretability. To complement each other, hybrid model-based DL approaches have gained interest.
For instance, Chennakeshava \textit{et al.}\ \cite{Chennakeshava:ius2020} solve a model-based plane-wave compounding problem by unfolding a classical optimization algorithm. 
Zhang \textit{et al.}~\cite{ZHANG2021} propose a self-supervised beamforming approach enforcing explicit prior assumptions on the reconstruction through the loss function. 
Our work falls within this model-based DL family of approaches, but contrary to previous methods which are deterministic, we focus on stochastic methods based on the recent success of 
Diffusion Models ~\cite{ho_denoising_2020,nichol_improved_2021,dhariwal_diffusion_2021}, renowned for image synthesis.

Instead of conventionally training a ``task-specific" neural network, diffusion-based inverse problem solvers use pre-trained neural networks to encode only the prior knowledge~\cite{DDRM,PGDM,DPS,song_solving_2022,chung2022scoreMRI,zhang2023}.
In a recent work~\cite{zhang2023}, we adapted the Denoising Diffusion Restoration Models (DDRMs) framework~\cite{DDRM} to incorporate the physics and constraints of US imaging through an approximate direct model, and with a self-supervised fine-tuning of the unconditional diffusion model. 
\DM{Also} related, 
Asgariandehkordi~\textit{et al.} \cite{DenoDDPM} apply diffusion denoising without considering 
a
measurement model.  Stevens \textit{et al.}~\cite{dehaze} focus on the specific task of dehazing cardiovascular US images.

The stochastic nature of a diffusion model leads to varying posterior samples under different noise initializations. Studies have demonstrated that the variance of these samples is non-uniform and 
\DM{tends to delineate edges or uncertain regions of the reconstructed objects}
~\cite{horwitz2022conffusion,chung2022scoreMRI,dehaze}. 
\DM{
Unlike such uncertainty interpretation, this paper proposes a novel perspective of the diffusion variance as an estimator for US images, following the nature of the} multiplicative noise present in US reflectivity maps. The contributions of our work are:
\begin{enumerate} 
\item 
Proposing an empirical model characterizing the stochasticity of 
\DM{diffusion-based US recontructions.}

\item
Revealing that computing the variance of multiple diffusion reconstructed samples 
\DM{achieves higher SNR and contrast, and results in a despeckling without oversmoothing.}
\item
Confirming the superiority of the inverse-problem-informed diffusion reconstruction method over the denoising diffusion approach~\cite{DenoDDPM} through experiments on real data.
\end{enumerate}

\section{Variance of a Diffusion Restoration Model}

\subsection{Denoising Diffusion Restoration Models}
Diffusion Denoising Probabilistic Models (DDPMs) are a class of parameterized Markov chains used to generate synthetic images from noise~\cite{ho_denoising_2020,nichol_improved_2021,dhariwal_diffusion_2021}. These models employ a fixed forward diffusion process and a learned backward generation process. During training, the forward process gradually adds Gaussian noise with variance $\sigma_t^2$ ($t = 1,\ldots, T$) to a clean image $\xv_0$ until it becomes random noise $\xv_T$, while the backward process iteratively denoises the pure noise until the original clean image is reconstructed. Sampling involves simulating 
backward paths
\DM{composed of multiple steps}
using either stochastic differential equations (SDEs) or ordinary differential equations (ODEs) to generate synthetic clean images.

Within the realm of model-based deep learning, a 
\DM{recent}
focus is made on leveraging learned prior knowledge from diffusion models to address inverse problems~\cite{DDRM, DPS, PGDM,song_solving_2022,chung2022scoreMRI,zhang2023}. Unlike conventional task-specific learning approaches relying on paired datasets~\cite{perdios_cnn-based_2022,van_sloun_deep_2020}, the above algorithms operate as Markov Chains conditioned on measurements through an \textit{observation model}, allowing the utilization of pre-trained diffusion models across multiple tasks 
\DM{instead of}
task-specific training.

Assume
an observation model $\yv_d= \Hv_d \xv_d + \nv_d$, where $\yv_d$ denotes the measurements, $\Hv_d$ is the degradation operator, $\xv_d$ is the 
sought quantity,
and $\nv_d$ represents \textit{i.i.d.} Gaussian additive noise with standard deviation $\sigma_d$.
For such problems,
Denoising Diffusion Restoration Models (DDRMs)~\cite{DDRM} serve as robust diffusion-based solvers operating in the spectral space of $\Hv_d$\footnote{We use subscript $d$ to refer to the original equations of DDRM.}. To this end, DDRM leverages the Singular Value Decomposition (SVD): 
$\Hv_d=\Uv_d \Sb_d \Vv_d\T$ with $\Sb_d=\Diag\left(s_1,\ldots,s_N\right)$,
to decouple the dependencies between the measurements, and cast the original observation model as a denoising problem:
$$
\overline{\yv}_d= \overline{\xv}_d + \overline{\nv}_d
$$
with $\overline{\yv}_d= \Sb_d^\dag\Uv_d\T\yv_d$, $ \overline{\xv}_d=\Vv_d\T\xv_d$, and $\overline{\nv}_d=\Sb_d^\dag\Uv_d\T \nv_d$, where $\Sb_d^\dag$ is the generalized inverse of $\Sb_d$.

The ODE-based sampling in DDRM enables fast restoration by covering multiple diffusion steps at once, \emph{i.e.}, directly transitioning from $\overline{\xv}_t$ to $\overline{\xv}_{t-k}$, where $k\geqslant 1$ denotes the number of skipped diffusion steps~\cite{DDIM}. 
In practice, the value of $\overline{\xv}_{t-k}$ is determined by linearly combining $\overline{\xv}_t$, the transformed measurements $\overline{\yv}_d$, the transformed current prediction $\overline{\xv}_{\theta,t}$, and random noise using coefficients $A$, $B$, $C$, and $D$. These coefficients satisfy conditions related to noise and signal, specifically $(A\sigma_t)^2 + (B\sigma_d/s_i)^2 + D^2 = {\sigma_{t-k}}^2$ for the noise and $A+B+C = 1$ for the signal. The adjustment of these coefficients is governed by two hyperparameters, providing flexibility in the restoration process. After $N_\text{it}$ denoising iterations, the final restored image $\xv_0$ is obtained as $\Vv_d\overline{\xv}_{0}$. 

\subsection{DRUS Variance Imaging}
Having introduced the general principle of DDRM, this section introduces the adaptation of this framework to integrate a direct model describing the construction of an US image, incorporating both additive and multiplicative noise.

For a given 
\DM{imaged}
area,
we denote $p(\rv)$ the echogenicity at position $\rv\in\Omega$, 
which can be modeled as piecewise smooth. The tissue reflectivity can then be expressed as:
\begin{equation}
    o(\rv)=m(\rv)p(\rv),
    \label{Equ: model_reflectivity}
\end{equation}
where $m$ refers to the standard Gaussian multiplicative noise~\cite{speckleModel2007Ng,perdios_cnn-based_2022}. 
\DM{To capture measurements from $p$,} US imaging uses a transducer array with $L$ elements, each emitting a pulse 
and receiving an echo signal. 
We consider the first-order Born approximation and no absorption within tissues to model the ultrasonic transmission-reception process as a linear time-invariant system.

The US wave emitted by the $i$th element passes through the object domain $\Omega$ and is received by the $j$th element. The received radio-frequency (RF) signal at time $t$ 
\DM{is then:}
\begin{equation}
    y_{i, j}(t) = \int_{\rv \in \Omega}h(t-\tau_{i,j}(\rv)) m(\rv)p(\rv) \mathrm{d} \rv + n_{j}(t),
\label{Equ: model_continuous}
\end{equation}
where $n_{j}$ represents the noise for the $j$th receiving element, $h$ is a kernel resulting from the convolution of the emitted excitation pulse $h_e$ and the transducer impulse response $h_t$.

For single plane-wave (PW) US imaging, the discretized model with $N$ observation positions and $K$ time samples for all $L$ receiving elements can then be written as: 
\begin{equation}
    \yv=\Hv(\mv\odot\pv) + \nv,
    \label{Equ: model_discret}
\end{equation}
where 
$\yv=[\yv_1\T,...,\yv_L\T]\T\in \R^{KL\times 1}$, $\Hv\in \R^{KL\times N}$, $\mv\odot\pv$ is the componentwise product of \mv by \pv, both vectors being in $\R^{N\times 1}$,
and $\nv=[\nv_1\T,...,\nv_L\T]\T\in \R^{KL\times 1}$ includes electronic noise and model error. For simplicity, $\nv$ is assumed white Gaussian with standard deviation $\gamma$, which is reasonable for the plane wave transmission~\cite{iMAP}. 

In the domain of medical US image reconstruction, some studies \cite{ZHANG2021,zhang2023} focus on restoring the reflectivity map, while others \cite{ADMSS,usDespeckle2022Lee} opt for reconstructing the echogenicity map.
Our previous work, DRUS (Diffusion Reconstruction of US images)~\cite{zhang2023}, falls into the former category. It projects the data in ~\eqref{Equ: model_discret} using a beamforming matrix\footnote{A weighted matched filter matrix of $\Hv$.} $\Bv \in \R^{N\times KL}$:
\begin{equation}
   \Bv\yv=\Bv\Hv\ov + \Bv\nv,
   \label{Equ: model_BH}
\end{equation}
and employs DDRM to estimate \DM{the reflectivity} \ov. The current paper explores a new application of DRUS, extending it to estimate US echogenicity \DM{\pv} maps.

Diffusion sampling in DRUS 
\DM{can yield}
varied $\hat\ov$ due to stochasticity. Research has shown that 
the variance of diffusion posterior samples under additive noise tends to highlight edges of reconstructed objects~\cite{horwitz2022conffusion, chung2022scoreMRI, dehaze}. However, due to the multiplicative noise inherent in US, the amplitudes of \ov follow zero-mean Gaussian distributions with standard deviation depending on the intensity of \pv, affecting neighboring pixel differences and thus diffusion variance. Therefore, the variance is intuitively linked to \pv, and we introduce the following 
model to characterize this property: 
\begin{align}
\hat\ov_c = \mv\odot\pv + \pv^\beta \odot\Gv_c,
\end{align}
where $\hat\ov_c$ represents the $c$th DRUS sample, $\Gv_c$ follows a standard normal distribution to account for generative stochasticity, and $\beta$ is an empirical parameter. It can be easily checked that $\ED[\hat\ov_c] = \mv\odot\pv$ and $\mathrm{Var}[\hat\ov_c] = \pv^{2\beta}$. Consequently, we define DRUSvar as an US echogenicity map estimator:
\begin{align}
\hat\pv_\text{DRUSvar} =  \left( \frac{1}{C-1}\sum_{c=1}^{C}\left|\hat\ov_c-\hat\ov_\text{DRUSmean}\right|^2 \right)^{\frac1{2\beta}},
\end{align}
where $\hat\ov_\text{DRUSmean} =  \frac{1}{C}\sum_{c=1}^{C}\hat\ov_c$ 
\DM{is hereafter named} 
DRUSmean. 
In our experiments, $\beta=0.5$ (used in this paper) empirically yielded the most favorable results.



\section{Numerical analysis}\label{sec: NumericalAnalysis}
\begin{figure}[h]
\centering
\includegraphics[width=0.45\linewidth]{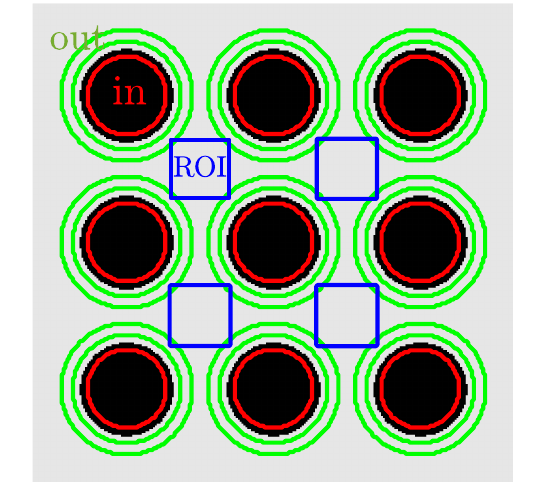}
\includegraphics[width=0.45\linewidth]{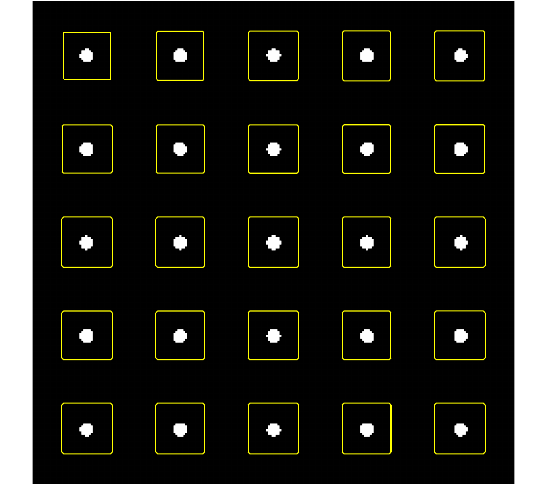}
\caption{Echogenicity maps of the 
\DM{synthetic occlusion (left) and scatterer (right) phantoms}. 
Metrics are calculated within the colored boundary regions.
} 
\label{Fig: numericalPhan}
\end{figure}

We focus on reconstructing US images from a single plane wave (PW). 
This section 
demonstrates the feasibility of using DRUSvar as an US echogenicity map estimator \DM{under a controlled}
numerical analysis on two synthetic phantoms.
As illustrated in Fig.~\ref{Fig: numericalPhan}, the occlusion phantom comprises 9 anechoic regions, while the second one contains 25 scatterers. For each phantom, we synthesize 9 reflectivity maps with various realizations of multiplicative noise, following \eqref{Equ: model_reflectivity}.

\DM{
To reproduce the behaviour of DRUSvar under a minimal observation model and emphasize on its pertinence in presence of multiplicative noise, we employ}
two 1-D convolution kernels instead of the full degradation matrix $\Bv\Hv$ in \eqref{Equ: model_BH}. Specifically, the lateral kernel is a Gaussian with a standard deviation of 0.17 mm, while the axial kernel is a cosine-modulated Gaussian with the same standard deviation to mimic a realistic US pulse-echo response (i.e., $h$ in \eqref{Equ: model_continuous}).

The restoration process employs a US fine-tuned diffusion model at a resolution of $256\times256$. The original open-source diffusion model
(Downloaded from \href{https://github.com/openai/guided-diffusion
}{https://github.com/openai/guided-diffusion})
was trained on ImageNet~\cite{ILSVRC15}. For fine-tuning, a dataset consisting of 2339 high-quality \textit{in vitro} images was acquired from a TPAC Pioneer machine using the CIRS 040GSE phantom. 

We conduct qualitative and quantitative evaluations on the above phantoms, considering various additive noise levels $std(\nv)$. All diffusion samples underwent $N_\text{it}=50$ iteration steps and the DRUSmean/var images were constructed with $C=10$ samples.
The occlusion phantom is used to evaluate the generalized Contrast to Noise Ratio (gCNR)~\cite{gCNR} and the Signal to Noise Ratio (SNR), defined as:
$$
\mathrm{gCNR}=1-\int_{-\infty}^{\infty} \min \left\{g_{\text {in}}(v), g_{\text {out}}(v)\right\} dv,
$$
$\textit{SNR} = \mu_\text{ROI} / \sigma_\text{ROI}$, respectively,
where the subscripts `in' and `out' indicate inside or outside the target regions, $v$ denotes the pixel values, $g$ refers to the histogram of pixels in each region, `ROI' refers to the region of interest, and $\mu$ and $\sigma$ denote the mean and the standard deviation respectively.
The spatial resolution evaluated using -6\,dB Full Width at Half Maximum (FWHM) is assessed on the scatterer phantom. A smaller FWHM value indicates a higher resolution.

\begin{figure}[ht]
\centering
\setlength{\tabcolsep}{1pt}
\small
\includegraphics[width=0.75\linewidth]{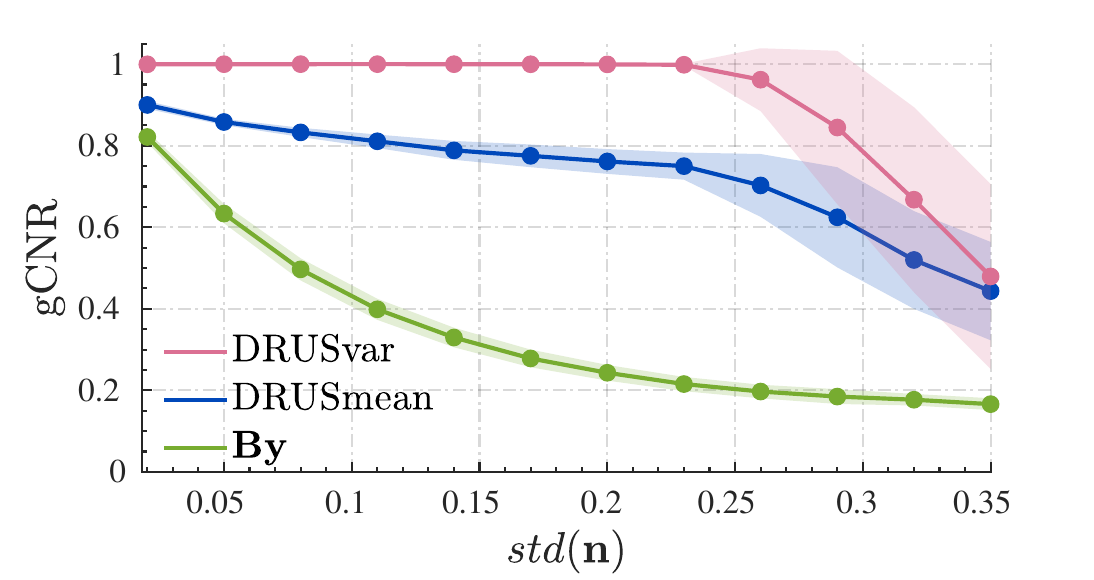}

\includegraphics[width=0.75\linewidth]{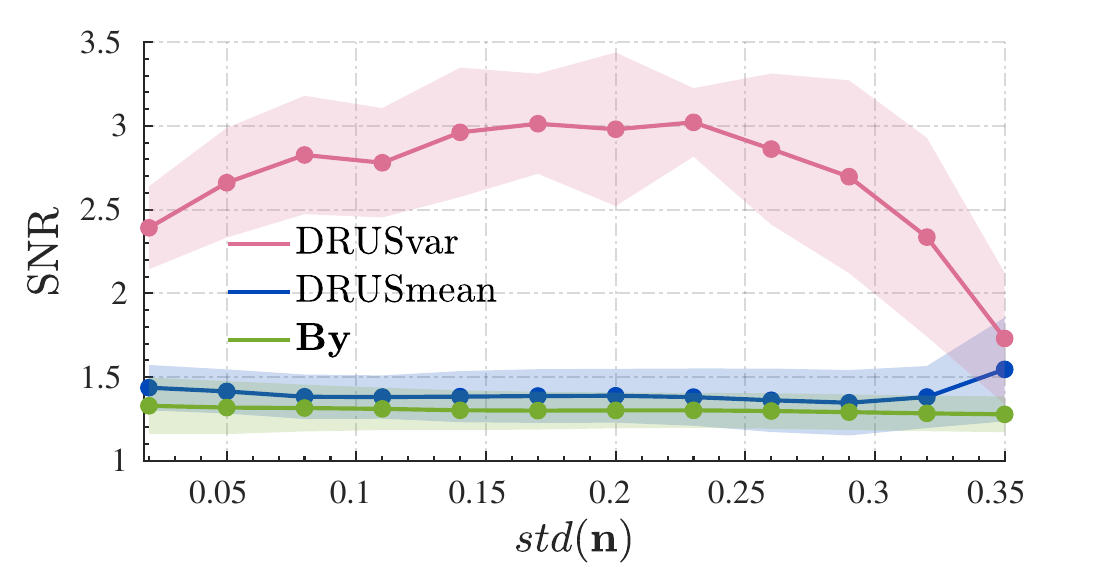}
\medskip

\begin{tabular}{rcc}
\cc{$\Bv\yv$}
&\figc[width=0.4\linewidth]{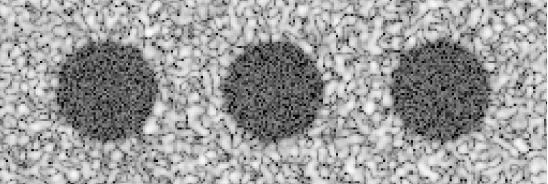}
&\figc[width=0.4\linewidth]{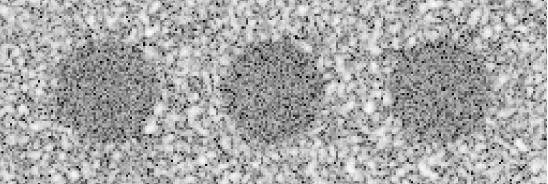}
\\
{\renewcommand{\arraystretch}{.8}
\begin{tabular}{c}
DRUS\\mean\end{tabular}}
&\figc[width=0.4\linewidth]{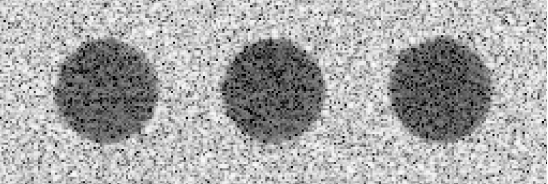}
&\figc[width=0.4\linewidth]{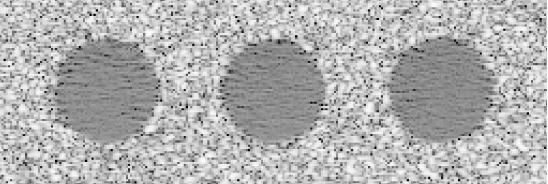}
\\
{\renewcommand{\arraystretch}{.8}
\begin{tabular}{c}
DRUS\\var\end{tabular}}
&\figc[width=0.4\linewidth]{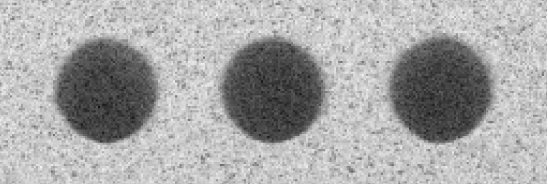}
&\figc[width=0.4\linewidth]{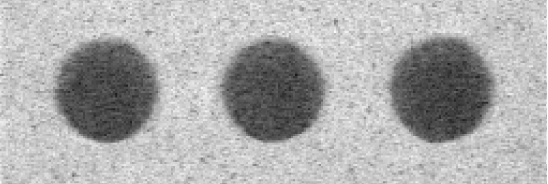}
\end{tabular}
\caption{Quantitative and qualitative comparison of the synthetic occlusion phantom-based images under varying levels of additive noise. 
Images are in decibels with a dynamic range [-60,0]. 
\DM{Left $std(\nv)=0.02$, right $std(\nv)=0.08$. }
} 
\label{Fig: cysts}
\end{figure}

\begin{figure}[ht]
\centering
\small
\includegraphics[width=0.75\linewidth]{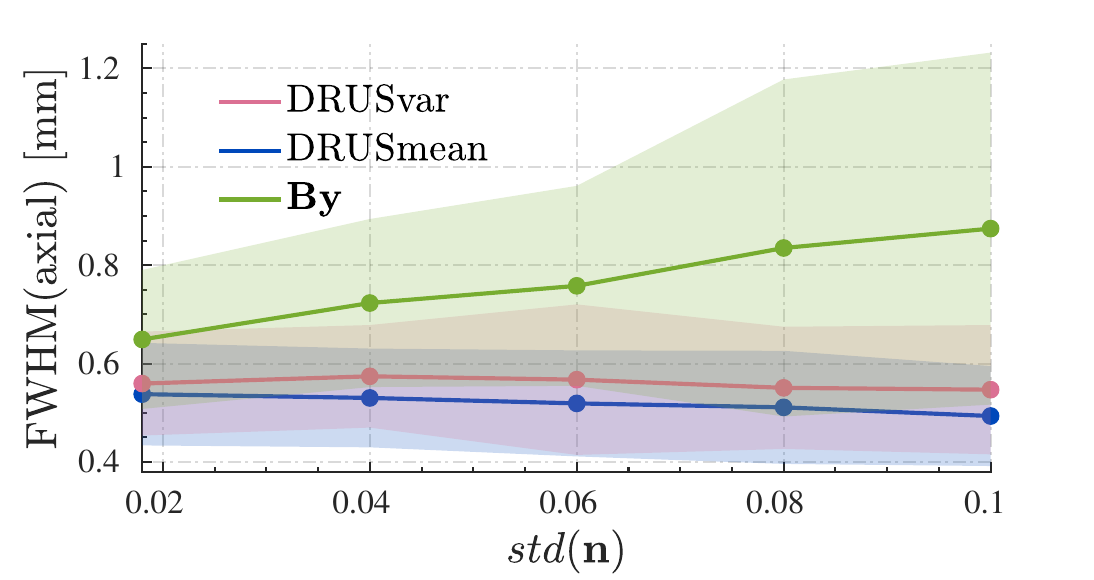}
\includegraphics[width=0.75\linewidth]{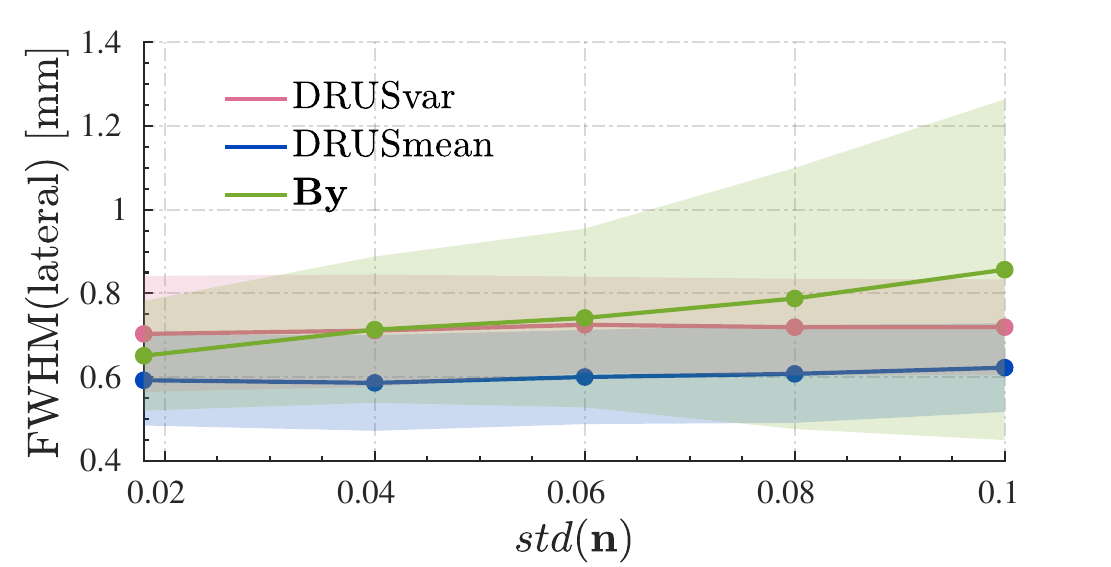}
\medskip

\setlength{\tabcolsep}{1pt}
\begin{tabular}{rcc}
\cc{$\Bv\yv$}
&\figc[width=0.4\linewidth]{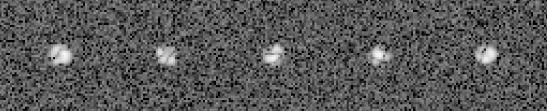}
&\figc[width=0.4\linewidth]{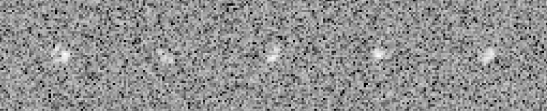}
\\
{\renewcommand{\arraystretch}{.8}
\begin{tabular}{c}
DRUS\\mean\end{tabular}}
&\figc[width=0.4\linewidth]{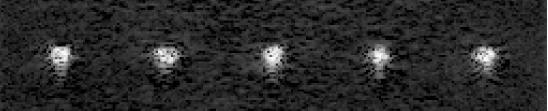}
&\figc[width=0.4\linewidth]{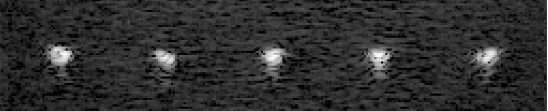}
\\
{\renewcommand{\arraystretch}{.8}
\begin{tabular}{c}
DRUS\\var\end{tabular}}
&\figc[width=0.4\linewidth]{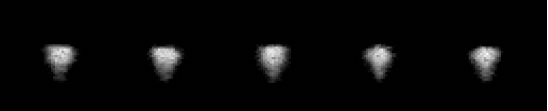}
&\figc[width=0.4\linewidth]{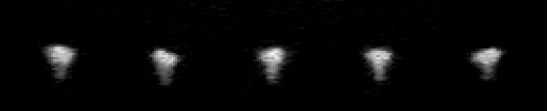}
\end{tabular}

\caption{Quantitative and qualitative comparison of the synthetic scatterer phantom-based images under varying levels of additive noise. Images are in decibels with a dynamic range [-60,0]. Left $std(\nv)=0.018$, right $std(\nv)=0.1$. }
\label{Fig: scatterers}
\end{figure}

The results 
in Figs.~\ref{Fig: cysts} (occlusion) and~\ref{Fig: scatterers} (scatterer) comprise the quantitative scores and the qualitative images. The means and the standard deviations of the scores were calculated across all target regions from the 9 synthetic reflectivity maps. 
Fig.~\ref{Fig: cysts} demonstrates the significant superiority of DRUSvar over DRUSmean \DM{and $\Bv \yv$} in terms of gCNR, SNR, and visualization. While DRUSvar exhibits slightly inferior performance compared to DRUSmean in terms of spatial resolution (Fig.~\ref{Fig: scatterers}), it remains stable even with large \DM{levels of} additive noise.

\section{Real data results}\label{sec:Realdata}

We conducted quantitative and qualitative evaluations of our approach based on the publicly available experimental datasets from the Plane Wave Imaging Challenge in Medical UltraSound (PICMUS)~\cite{PICMUS}. The datasets were obtained with a 128-element L11–4v linear-array transducer. The transmit pulse has a central frequency of 5.208 MHz and a bandwidth ratio (BWR) of 67\%. The sampling rate is 20.8 MHz. We name the 2 \textit{in vitro} datasets \textit{EC (Experimental Contrast)} and \textit{ER (Experimental Resolution)}, and the 2 \textit{in vivo} datasets \textit{CC (Carotid Cross-sectional)} and \textit{CL (Carotid Longitudinal)}.

The linear inverse problem model in \eqref{Equ: model_BH} is leveraged in DRUSmean/var. The construction of the matrix $\Hv$ necessitates the channel data acquisition parameters, the field of view, and the image resolution. The data acquisition parameters came from the PICMUS setup, the field of view spans from -18 mm to 18 mm in width and from 10 mm to 46 mm in depth, with the origin located at the transducer center. The image resolution was fixed at $256 \times 256$. The construction of the beamforming matrix $\Bv$ relies on the same parameters as $\Hv$, and the receive apodization weights defined by a window of Tukey0.25 and an \textit{f-number} of 1.4.

\begin{figure*}[ht]
\centering
\includegraphics[width=0.9\linewidth]{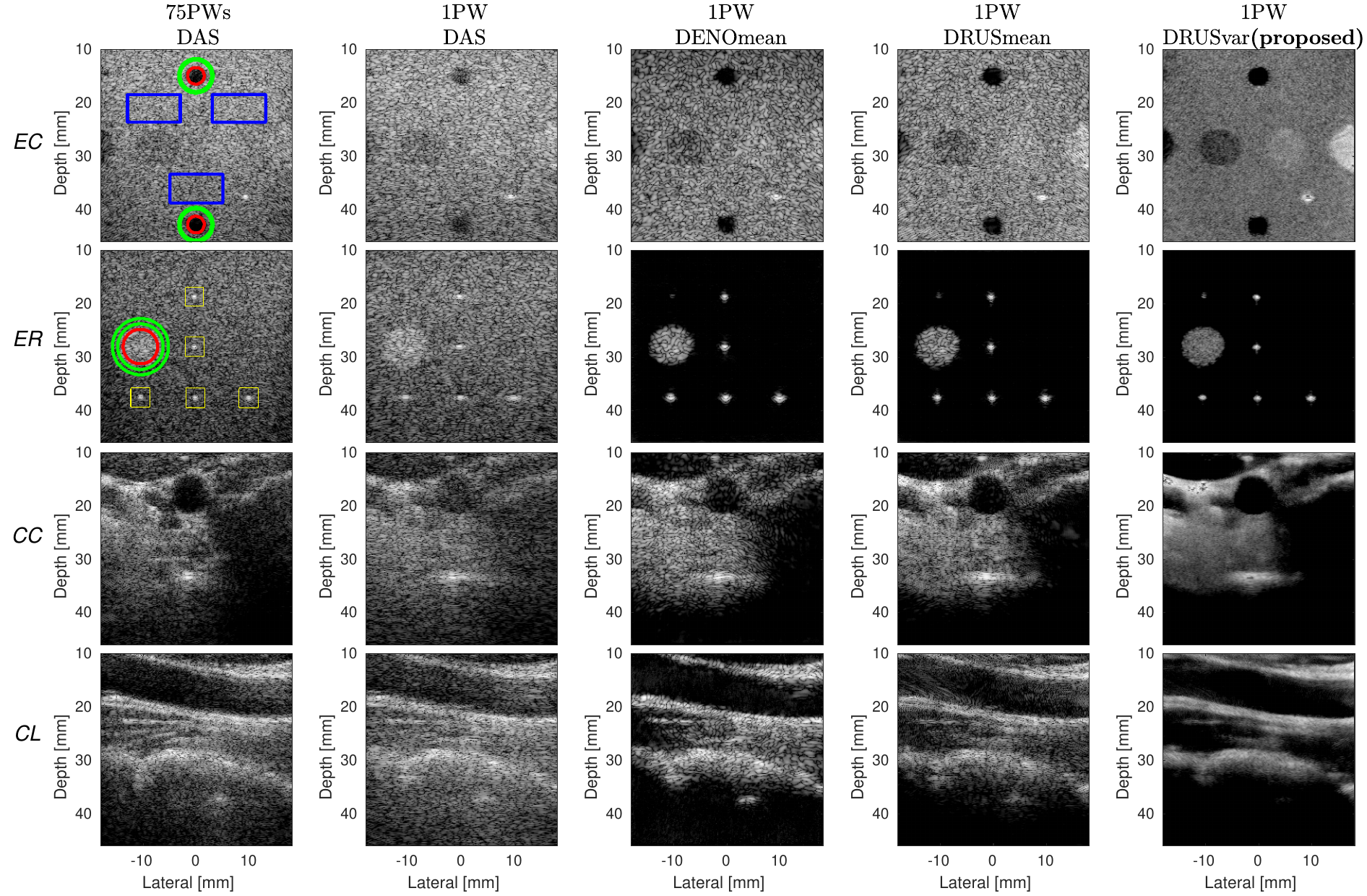}
\caption{Comparison of reconstructed images
on the PICMUS \textit{in vitro} (row [1-2]) and \textit{in vivo} (row [3-4]) datasets. All images are in decibels with a dynamic range [-60,0]. The colored boundaries outline the regions where the \textit{in vitro} evaluation metrics are calculated.} 
\label{Fig: Picmus}
\end{figure*}

\begin{figure}[ht]
\centering
\includegraphics[width=0.8\linewidth]{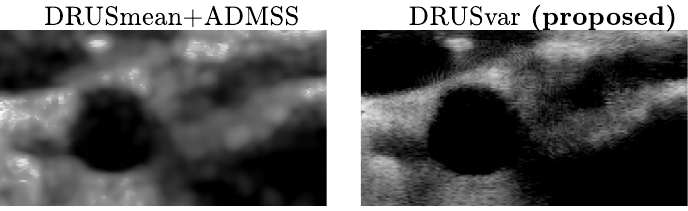}
\caption{Visual comparison of despeckled images on the \textit{CC} dataset, in decibels [-60,0] dynamic range.} 
\label{Fig: edgesCompare}
\end{figure}

The same diffusion model and evaluation metrics as in Sect.~\ref{sec: NumericalAnalysis} are employed for \textit{EC} and \textit{ER}. \textit{CC} and \textit{CL} use a diffusion model fine-tuned on 1012 home-made \textit{in vivo} images. $N_\text{it}=50$ and $C=10$. 
Quantitative results are presented in Table~\ref{Tab: vitroPicmus}. Fig.~\ref{Fig: Picmus} and \ref{Fig: edgesCompare} depict the qualitative results.
In addition to comparing DRUSvar with DAS (1PW and 75 PWs) and DRUSmean, we also include a comparison against DENOmean with 10 DENO~\cite{DenoDDPM} samples, a state-of-the-art technique that denoises US images with diffusion models without solving an inverse problem. Furthermore, for despeckling comparison, DRUSvar is visually contrasted with DRUSmean+ADMSS~\cite{ADMSS}\footnote{Anisotropic Diffusion with Memory based on Speckle Statistics for Ultrasound Images. Code at \href{https://fr.mathworks.com/matlabcentral/fileexchange/52988-anisotropic-diffusion-with-memory-based-on-speckle-statistics-for-ultrasound-images?s_tid=FX_rc2_behav}{MathWorks}.}, where ADMSS is an US despeckling method applied on beamformed images before log compression.

Table~\ref{Tab: vitroPicmus} shows that our approach, DRUSvar, significantly outperforms others in terms of contrast and SNR, while maintaining competitive spatial resolution. This is consistent with the numerical observations in Sect.~\ref{sec: NumericalAnalysis} and the qualitative results in Fig.~\ref{Fig: Picmus}. Additionally, Fig.~\ref{Fig: edgesCompare} demonstrates that DRUSvar mitigates the over-smoothing issue commonly seen in US despeckling methods like ADMSS~\cite{ADMSS}. However, for applications such as motion tracking, where speckle is useful, DRUSvar, being a despeckling method, is not suitable.

\begin{table}[ht]
\caption{Quantitative comparison
on the PICMUS \textit{in vitro} datasets. A and L denote axial and lateral directions respectively. Best values bolded, second-best underlined.}
\label{Tab: vitroPicmus}
{%
\centering
\setlength{\tabcolsep}{1pt} 
\begin{tabular}{cc@{\,}c|cccccc}
\hline
 &  &  & \multicolumn{2}{c}{DAS} & ~DENOmean~ & ~DRUSmean~ & DRUSvar \\
 &  &  & 75\,PWs & 1PW & 1PW\,\cite{DenoDDPM} & 1PW,\,\cite{zhang2023} & 1PW\,\textcolor{red}{(proposed)} \\ \hline
\multicolumn{1}{c@{~}|}{\multirow{2}{*}{\textit{EC}}} & \multicolumn{2}{c|}{gCNR$\uparrow$} & 0.95 & 0.87 & 0.95 & \underline{0.97} & \textbf{0.98} \\
\multicolumn{1}{c|}{} & \multicolumn{2}{c|}{SNR$\uparrow$} & 1.92 & \underline{1.97} & 1.93 & 1.87 & \textbf{3.03} \\ \hline

\multicolumn{1}{c@{~}|}{\multirow{3}{*}{\textit{ER}}} & \multicolumn{2}{c|}{gCNR$\uparrow$} & 0.77 & 0.69 & \underline{0.95} & \underline{0.95} & \textbf{1.00} \\
\multicolumn{1}{c|}{} & \multirow{2}{*}{\begin{tabular}[c]{@{}c@{}}FWHM\\ {[}mm{]}\end{tabular}} & A$\downarrow$ & 0.54 & 0.56 & \underline{0.31} & \textbf{0.24} & 0.34 \\
\multicolumn{1}{c|}{} &  & L$\downarrow$ & 0.56 & 0.87 & 0.64 & \underline{0.54} & \textbf{0.32} \\ \hline
\end{tabular}%
}
\end{table}

\section{Conclusion}

This paper proposes an empirical model to characterize the stochasticity of diffusion reconstruction of US images, and introduces a novel US imaging approach, DRUSvar. The proposed approach leverages the empirical variance of multiple DRUS~\cite{zhang2023} samples to mitigate speckle noise, thereby enhancing contrast and SNR, without causing over-smoothing. The feasibility and competitiveness of DRUSvar are demonstrated in the task of single plane-wave US image restoration, using both synthetic and real-world datasets.

\bibliographystyle{IEEEtran} 
\bibliography{IEEEabrv,references} 

\begin{thebibliography}{10}
\providecommand{\url}[1]{#1}
\csname url@samestyle\endcsname
\providecommand{\newblock}{\relax}
\providecommand{\bibinfo}[2]{#2}
\providecommand{\BIBentrySTDinterwordspacing}{\spaceskip=0pt\relax}
\providecommand{\BIBentryALTinterwordstretchfactor}{4}
\providecommand{\BIBentryALTinterwordspacing}{\spaceskip=\fontdimen2\font plus
\BIBentryALTinterwordstretchfactor\fontdimen3\font minus \fontdimen4\font\relax}
\providecommand{\BIBforeignlanguage}[2]{{%
\expandafter\ifx\csname l@#1\endcsname\relax
\typeout{** WARNING: IEEEtran.bst: No hyphenation pattern has been}%
\typeout{** loaded for the language `#1'. Using the pattern for}%
\typeout{** the default language instead.}%
\else
\language=\csname l@#1\endcsname
\fi
#2}}
\providecommand{\BIBdecl}{\relax}
\BIBdecl

\bibitem{DAS}
V.~Perrot, M.~Polichetti, F.~Varray, and D.~Garcia, ``So you think you can {DAS}?'' \emph{Ultrasonics}, vol. 111, p. 106309, 2021.

\bibitem{IPB_Ozkan}
E.~Ozkan, V.~Vishnevsky, and O.~Goksel, ``Inverse problem of ultrasound beamforming with sparsity constraints and regularization,'' \emph{IEEE Trans. Ultrason. Ferroelectr. Freq. Control}, vol.~65, pp. 356--365, 2018.

\bibitem{RED_USIPB}
S.~Goudarzi, A.~Basarab, and H.~Rivaz, ``Inverse problem of ultrasound beamforming with denoising-based regularized solutions,'' \emph{IEEE Trans. Ultrason. Ferroelectr. Freq. Control}, vol.~69, 2022.

\bibitem{perdios_cnn-based_2022}
D.~Perdios, M.~Vonlanthen, F.~Martinez, M.~Arditi, and J.-P. Thiran, ``{NN}-based image reconstruction method for ultrafast ultrasound imaging,'' \emph{IEEE Trans. Ultrason. Ferroelectr. Freq. Control}, vol.~69, pp. 1154--1168, 2022.

\bibitem{van_sloun_deep_2020}
R.~J.~G. van Sloun, R.~Cohen, and Y.~C. Eldar, ``Deep learning in ultrasound imaging,'' \emph{Proc. IEEE}, vol. 108, pp. 11--29, 2020.

\bibitem{Chennakeshava:ius2020}
N.~Chennakeshava, B.~Luijten, O.~Drori, M.~Mischi, Y.~C. Eldar, and R.~J.~G. van Sloun, ``High resolution plane wave compounding through deep proximal learning,'' in \emph{IEEE IUS}, 2020.

\bibitem{ZHANG2021}
J.~Zhang, Q.~He, Y.~Xiao, H.~Zheng, C.~Wang, and J.~Luo, ``Ultrasound image reconstruction from plane wave radio-frequency data by self-supervised deep neural network,'' \emph{Med. Image Anal.}, vol.~70, 2021.

\bibitem{ho_denoising_2020}
J.~Ho, A.~Jain, and P.~Abbeel, ``Denoising diffusion probabilistic models,'' \emph{NeurIPS}, vol.~33, pp. 6840--6851, 2020.

\bibitem{nichol_improved_2021}
A.~Q. Nichol and P.~Dhariwal, ``Improved denoising diffusion probabilistic models,'' in \emph{ICML}, 2021, pp. 8162--8171.

\bibitem{dhariwal_diffusion_2021}
P.~Dhariwal and A.~Nichol, ``Diffusion models beat {GANs} on image synthesis,'' \emph{NeurIPS}, vol.~34, pp. 8780--8794, 2021.

\bibitem{DDRM}
B.~Kawar, M.~Elad, S.~Ermon, and J.~Song, ``Denoising diffusion restoration models,'' \emph{NeurIPS}, vol.~35, pp. 23\,593--23\,606, 2022.

\bibitem{PGDM}
J.~Song, A.~Vahdat, M.~Mardani, and J.~Kautz, ``Pseudoinverse-guided diffusion models for inverse problems,'' in \emph{ICLR}, 2022.

\bibitem{DPS}
H.~Chung, J.~Kim, M.~T. Mccann, M.~L. Klasky, and J.~C. Ye, ``Diffusion posterior sampling for general noisy inverse problems,'' in \emph{ICLR}, 2022.

\bibitem{song_solving_2022}
Y.~Song, L.~Shen, L.~Xing, and S.~Ermon, ``Solving inverse problems in medical imaging with score-based generative models,'' in \emph{ICLR}, 2022.

\bibitem{chung2022scoreMRI}
H.~Chung and J.~C. Ye, ``Score-based diffusion models for accelerated {MRI},'' \emph{Med. Image Anal.}, vol.~80, p. 102479, 2022.

\bibitem{zhang2023}
Y.~Zhang, C.~Huneau, J.~Idier, and D.~Mateus, ``Ultrasound image reconstruction with denoising diffusion restoration models,'' in \emph{DGM4MICCAI}, 2023, pp. 193--203.

\bibitem{DenoDDPM}
H.~Asgariandehkordi, S.~Goudarzi, A.~Basarab, and H.~Rivaz, ``Deep ultrasound denoising using diffusion probabilistic models,'' in \emph{IEEE IUS}, 2023.

\bibitem{dehaze}
T.~S. Stevens, F.~C. Meral, J.~Yu, I.~Z. Apostolakis, J.-L. Robert, and R.~J. Van~Sloun, ``Dehazing ultrasound using diffusion models,'' \emph{IEEE Trans. Med. Imaging}, 2024.

\bibitem{horwitz2022conffusion}
E.~Horwitz and Y.~Hoshen, ``Conffusion: Confidence intervals for diffusion models,'' \emph{arXiv:2211.09795}, 2022.

\bibitem{DDIM}
J.~Song, C.~Meng, and S.~Ermon, ``Denoising diffusion implicit models,'' in \emph{ICLR}, 2021.

\bibitem{speckleModel2007Ng}
J.~Ng, R.~Prager, N.~Kingsbury, G.~Treece, and A.~Gee, ``Wavelet restoration of medical pulse-echo ultrasound images in an {EM} framework,'' \emph{IEEE Trans. Ultrason. Ferroelectr. Freq. Control}, vol.~54, 2007.

\bibitem{iMAP}
T.~Chernyakova, D.~Cohen, M.~Shoham, and Y.~C. Eldar, ``{iMAP} beamforming for high-quality high frame rate imaging,'' \emph{IEEE Trans. Ultrason. Ferroelectr. Freq. Control}, vol.~66, 2019.

\bibitem{ADMSS}
G.~Ramos-Llord{\'e}n, G.~Vegas-S{\'a}nchez-Ferrero, M.~Martin-Fernandez, C.~Alberola-L{\'o}pez, and S.~Aja-Fern{\'a}ndez, ``Anisotropic diffusion filter with memory based on speckle statistics for ultrasound images,'' \emph{IEEE Trans. Image Process.}, vol.~24, pp. 345--358, 2014.

\bibitem{usDespeckle2022Lee}
H.~Lee, M.~H. Lee, S.~Youn, K.~Lee, H.~M. Lew, and J.~Y. Hwang, ``Speckle reduction via deep content-aware image prior for precise breast tumor segmentation in an ultrasound image,'' \emph{IEEE Trans. Ultrason. Ferroelectr. Freq. Control}, vol.~69, pp. 2638--2650, 2022.

\bibitem{ILSVRC15}
O.~Russakovsky \emph{et~al.}, ``Imagenet large scale visual recognition challenge,'' \emph{Int. J. Comput. Vis.}, vol. 115, pp. 211--252, 2015.

\bibitem{gCNR}
A.~Rodriguez-Molares \emph{et~al.}, ``The generalized contrast-to-noise ratio: A formal definition for lesion detectability,'' \emph{IEEE Trans. Ultrason. Ferroelectr. Freq. Control}, vol.~67, pp. 745--759, 2020.

\bibitem{PICMUS}
H.~Liebgott, A.~Rodriguez-Molares, F.~Cervenansky, J.~Jensen, and O.~Bernard, ``Plane-wave imaging challenge in medical ultrasound,'' in \emph{IEEE IUS}, 2016.

\end{thebibliography}

\end{document}